\documentclass[multphys,vecphys]{svmult}

\usepackage{makeidx}     % allows index generation
\usepackage{graphicx}    % standard LaTeX graphics tool
\usepackage{multicol}    % used for the two-column index
\makeindex             % used for the subject index
\begin{document}

\title*{The Mass Function of Newly Formed Stars (Review)}
\author{Lynne A. Hillenbrand\inst{1}}
\institute{California Institute of Technology;
MS 105-24; Pasadena, CA 91125 
\texttt{lah@astro.caltech.edu}}
\maketitle

\section{Introduction}

The topic of the stellar ``original mass function" has a nearly 50 year history,
dating to the publication in 1955 of Salpeter's seminal paper.  In this review
I discuss the many more recent results that have emerged
on the initial mass function (IMF), as it is now called,
from studies over the last decade of resolved populations 
in star forming regions and young open clusters.

The spiral structure of our Galaxy can be traced over many kilo parsecs near 
the plane ($b < 5^\circ$) via indicators of imminent, ongoing, or recent 
star formation such as molecular gas, HII regions, and OB stars. 
Locally, however, many of these same diagnostics can be found
far above the Galactic plane.  
The ring of local star formation, historically termed the Gould Belt 
(Gould 1879), is inclined by $\sim20^\circ$ with respect to the plane, 
and extends as close as 150 pc at its nearest and 750 pc 
at its furthest.  Towards the outer part of the Galaxy along this ring 
are the Perseus, Orion, and Monoceros clouds, crossing
the plane in Cygnus/Cepheus, and towards the inner Galaxy the 
Ophiuchus, Scorpius, Centaurus, Lupus, and Chamaeleon clouds
again crossing the plane in Vela.  Notably the famed
Taurus-Auriga star forming region is not part of the Gould's Belt but rather 
located in the vicinity of the inferred center of the expanding ring, near
the PerOB3 groups and dispersed Cas-Tau association. 
The Gould's Belt phenomenon recently has been 
repopularized based on the nearby young ($<$100 Myr) field stellar population 
uncovered by the Hipparcos and ROSAT all-sky surveys, as well as from new
appreciation of the spatial distribution of young neutron stars, pulsars, and
gamma ray sources.
The fortuitous geometry of local star formation enables us to study stellar 
populations relatively free of the 
foreground / background contamination and severe crowding and confusion which dominate
investigations of more distant Galactic plane star formation.  

Looking in detail at an individual star-forming molecular cloud one finds that the stellar
distribution is not random across the face of the cloud, but rather that young stars
tend to congregate into small groups or aggregates and denser clusters.  In Taurus,
for example, there are distinct regions of near-simultaneous aggregated
star formation in addition to similarly aged stars sparsely spread across 
the cloud (e.g. Hartmann et al 2001). 
In Orion, which is a more massive molecular complex by a factor of $\sim$10, 
the clusters (such as the Orion Nebula Cluster and NGC 2024)
are denser but a similar
dispersed population is still present (e.g. Strom, Strom \& Merrill 1993; 
Carpenter 2000).
Summarizing the global clustering characteristics of the local regions of
star formation, cluster densities range from unbound aggregates
or groupings of tens of stars over a 0.1-0.5 pc radius ($<$10$^2$ stars/pc$^3$)
to possibly bound clusters containing hundreds to thousands of stars 
over a 0.5-5 pc radius (10$^3$-10$^5$ stars/pc$^3$).
Furthermore, higher mass stars are associated with denser though not 
necessarily more populous clusters (Hillenbrand 1995; Testi et al. 1997). 
High-mass and low-mass stars do form in the same places and at the same times;
drastically different star formation mechanisms, e.g. for the production of 
massive stars, apparently are not required.  
Given the wide dynamic range in mass of stars 
and indeed sub-stellar mass objects in star forming regions, such arenas 
are natural targets for studies of the IMF.  

Investigation of the detailed form of the stellar/sub-stellar IMF in young
clusters is motivated by the fact that the distribution of masses is one 
of the most basic characteristics of a stellar system.  For a single region or
cluster, a successful theory of star formation should predict 
the salient features of the IMF such as minimum / maximum masses 
and the location / width of any peak in the distribution.   
A related predisposition concerns the universality of the IMF.
Molecular cloud fragmentation and star formation processes could lead to an IMF that
depends on variables such as time, location, metallicity, local stellar density,
cloud density or temperature, the pre-existence of massive stars, or whether
star formation is stochastic or triggered (e.g. via external cloud compression).
The detailed shape of the IMF is also important for several utilitarian 
reasons, particularly if the form is a universal one, including deduction of
the past star formation rate in the Galaxy (increasing,
decreasing, or constant in time), interpretation of integrated light from
distant regions in our Galaxy or other galaxies, and 
constraining the fraction of baryonic mass tied up in low 
luminosity objects.

In advance of summarizing the large body
of work on the IMF in star forming regions and young open clusters
it is useful to call attention to
pioneers of field star IMF studies, including Salpeter (1955), Miller \& Scalo (1979),
Scalo (1986), Rana (1987), Kroupa, Tout \& Gilmore (1993), Kroupa (2001).  The
field IMF near the stellar/sub-stellar boundary has been explored by 
Reid et al. (1999) and Burgasser et al. (2001).  High mass populations 
largely in clusters are summarized by Massey (1995).  These papers 
include many analysis steps in transforming observational 
data to masses.  First are the conversions of photometric 
and spectroscopic data to absolute magnitudes or luminosities and 
effective temperatures, and then the
corrections for incompleteness, star formation histories, stellar evolution,
and in some cases effects such as binarity.  Useful are recent review 
articles on the IMF including those by lead authors
Kroupa, Meyer, Elmegreen, Scalo, and Larson.  

What we know from these field star
studies is that the distribution of masses rises as 
dN/dM $\propto$ M$^{-\alpha}$ with the exponent $\alpha$ changing from 
$\sim$2.3 at the highest masses (nearly identical to the original Salpeter result which 
probed masses 0.3-16 M$_\odot$) to $\sim$0.3 near and across the sub-stellar
boundary.  While the slope does flatten towards lower masses, it should be emphasized
that the IMF extends in a continuous manner from the stellar domain well into
to the brown dwarf regime.  As found both in the field and in star forming 
regions, there is nothing special about the hydrogen-burning limit in terms 
of the production rates of gravitationally bound objects from the ISM.  

The various field star IMF's that have been proposed do have different implications for
the average stellar mass and the mass fraction in low- / intermediate- / high-mass objects.  For example, considering
a stellar population which extends from 0.1-120 M$_\odot$, a Salpeter (1955) IMF predicts 
that 96\% of the stars and 60\% of the mass are found at masses $<1 M_\odot$, with a mean
stellar mass of 0.35 M$_\odot$.  The log-normal form of the Miller-Scalo (1979) IMF,
on the other hand, predicts that 78\% of the stars and 31\% of the mass are found at
masses $<1 M_\odot$ with a mean stellar mass of 0.89 M$_\odot$, while the multiple 
power law Kroupa (2001) prediction is that 90\% of the stars and 45\% of the mass 
are found at masses $<1 M_\odot$, with a mean stellar mass of 0.80 M$_\odot$.  Extending
the lower mass limit of the population through the brown dwarf domain to 0.01 M$_\odot$
implies for the Kroupa IMF that 37\% of objects and 4\% of the mass are found
in brown dwarfs, and that the mean mass per object (star / brown dwarf)
drops to 0.38 M$_\odot$.  An additional way in which to think of the IMF as a
probability distribution is to ask how many stars are expected in a cluster before a
star of given mass is present.  Again using the Kroupa multiple power-law slope one
finds that 16 stars are needed before a 1 M$_\odot$ star is expected, 245 before the
first 8 M$_\odot$ star is expected, and 8274 before a 120 M$_\odot$
object is expected.  Drastically different numbers result from the various
proposed IMF's and the implications are not inconsequential for calculations
of quantities such as the number of ionizing photons or the number of
proto-supernovae in a cluster.

\section{Advantages and Disadvantages of Studying the IMF in 
Star Forming Regions and Young Open Clusters}

In contrast to the 
field star studies, investigations of youthful environments 
have the advantage
that all stars formed in the vicinity are still present; none have evolved 
into stellar remnants and
none have escaped the gravitational grasp of the molecular cloud and/or dense
stellar cluster.  
Thus what is measured is not only the present day mass function
but the initial mass function as well, with no corrections needed.  
The IMF can thus be measured over a large dynamic range, ideally from
$>$100 M$_\odot$ down to $<$0.08 M$_\odot$ in a single region. 
Importantly, young low mass stars and brown dwarfs are several (3-5) orders of
magnitude more luminous at 1-10 Myr of age than they are on the main sequence,
compressing the mass-luminosity relationship.  Further,
forming sub-stellar populations are much warmer than those in the solar neighborhood field
which have been cooling for many Gyr. At 1 Myr of age any object with spectral type later
than M6-M7 (i.e. T $<$ 3000 K ) is a brown dwarf ($<$0.08 M$_\odot \approx$ 80 M$_{Jupiter}$) 
and by mid-L has mass $<$10 M$_{Jupiter}$ according to most current pre-main 
sequence contraction models.   
Another advantage of young cluster populations for IMF studies is that
individual regions are co-eval, equi-distant, and have the same metallicity.  
Finally, presence of molecular gas can screen against contaminating background 
stars, making membership easier to assume.  
Results from one region can be compared 
directly to those from another with perhaps different
total mass, stellar density, or metallicity, and also to those from the solar
neighborhood field which represents a mix of star formation episodes occurring 
across kpc of space and Gyr of time in our Galaxy.

However, there are also several disadvantages to working in youthful environments.
These include the poorly calibrated evolution of the mass-luminosity relationship
through the pre-main sequence stage, the presence of large and spatially variable
extinction in star-forming regions, the possibility of photometric excesses due
to processes associated with circumstellar (accretion) disks, the spatial crowding
and nebulosity which complicate observations, and finally the lack of truly 
superb statistics for individual regions.

\section{Methodology for IMF Study of Young Stars}

With the above benefits and limitations in mind,
various techniques have been developed in attempting to measure the IMF 
in star-forming regions and young open clusters.  These include
the use of luminosity functions and a model of the star formation history 
(e.g. Zinnecker, Rieke, Comeron, Wilking, Strom, Lada, Alves, Muench), employment 
of color-magnitude diagrams and a photometric de-reddening technique as well as
a model of the star formation history (e.g. Meyer, Carpenter, McCaughrean,
Figer), and finally the construction of Hertzsprung-Russell
diagrams which require both spectroscopy and photometry (e.g. Herbig, Strom, 
Hillenbrand, Brown, Luhman, Preibisch, Rebull).   
All of these methods must assume a set of theoretical mass-luminosity
relationships and translate them into the needed format for comparison
to the data.

The spectroscopic techniques are the most robust, but consequently require the most
observational and analytical effort.  Feasibility with large telescopes (8-10 m)
extends well into the sub-stellar regime for the more proximate ($<$500 pc) 
regions of recent star formation, but the technique is applicable only 
to the upper end of the IMF in more distant ($>$1-2 kpc) regions.   
Spectroscopy is usually performed in the V, I, J, or K bands (0.5-2 $\mu$m) 
and enables the photospheric temperature to be estimated through adoption of 
a spectral-type to effective temperature relationship.  Spectral types are
derived from ratios of specific atomic lines or molecular bands with the most diagnostic
features generally found in the spectral regime closest to the peak wavelength of the
stellar blackbody.  The calibrations to effective temperature are established 
through fundamental techniques such as the infrared flux method, 
measured diameters, or for intrinsically less luminous objects via theoretical
modelling of stellar atmospheres.  Once a spectral type is determined 
an intrinsic color scale is used to compare observed to expected (given 
the spectral type) colors and 
thereby derive the color excess.  This can be converted using an adopted extinction law
into equivalent V-band extinction.   De-reddened photometry is then combined with a
bolometric correction scale to produce bolometric luminosities.   Finally, the derived
effective temperature and bolometric luminosity are compared to those predicted from
pre-main sequence evolutionary models to infer stellar masses and ages.

It should not go without mention that the pre-main sequence evolutionary models 
(e.g. those by D'Antona \& Mazzitelli 1997; Swenson et al. 1994;  Baraffe et al.
1998; Siess et al. 2000; Palla \& Stahler 2000; Yi et al. 2003)
remain poorly calibrated.  Variations between these predictions can be as much as
factors of 2 in mass and factors of 3 in age.
In most cases such models combined with quantities
derived from observational data such as Hertzsprung-Russell diagrams, color-magnitude
diagrams, or luminosity functions are the only means available for inferring stellar
masses.  However, fundamentally determined masses are those calculated 
from observations of dynamical
systems such as eclipsing double-line spectroscopic binaries or, to lower precision,
systems surrounded by circumstellar disks whose keplerian rotation can be measured.
Comparisons between the (limited) available dynamical masses and pre-main
sequence evolutionary tracks demonstrate that most existing 
sets of tracks systematically under-predict low-mass pre-main sequence masses 
by up to 20-30\% (Hillenbrand \& White 2004).

\section{IMF Results in Youthful Regions}

Before summarizing and synthesizing the results on stellar/sub-stellar IMFs
in star-forming regions and young open clusters over
the past decade, I call attention to a few recent results 
involving my own collaborators as examples of the sort of work being done.
First at the high-mass end, Slesnick, Hillenbrand, \& Massey (2002) performed 
a UBV photometric $+$ BV-band spectroscopy survey of the H/$\chi$ Per region, 
finding a typical Salpeter-like IMF for each of the two clusters 
($\Gamma=-1.36\pm0.20$ and $\Gamma=-1.25\pm0.23$) over
the mass range 4-40 M$_\odot$ with $\sim$40 past supernova expected.   At the 
other end of the mass spectrum, Slesnick, Hillenbrand, \& Carpenter (2004) 
have studied the sub-stellar domain of the Orion Nebula Cluster using 
JHK photometry and J/K spectra, confirming the turnover suspected by
several authors based on analyses at higher masses 
(Hillenbrand 1997; Luhman 2000; Muench et al 2002).     
Depending on the behavior of the mass-luminosity-temperature relationships --
which are still highly uncertain at such low masses -- there may even be
a flattening or potential rise towards the lowest mass brown dwarfs.

Generalizing now to the local set of star forming regions 
and young open clusters, it is often the case that incompleteness
and overall poor statistics preclude robust measurement
of the stellar/sub-stellar IMF.  Assembly of existing literature demonstrates
that close to 5000 stars in 24 regions can be located on the HR diagram and
compared to theoretical evolutionary tracks.  However, only a few
regions are reasonably complete into the brown dwarf 
regime, namely the Orion Nebula Cluster, IC 348, and the Pleiades.  These
regions compare well to a log-normal type (Miller-Scalo 1979)
mass function with acceptable statistical and systematic (due to methodology)
variation.

To provide a meaningful assessment of the IMF, observers need to measure its shape
in detail, including the dynamic range, peak, and width of the peak.  Attempts at
functional fits are welcome.  At present, however, most authors simply force 
single power laws through their data and quote the resulting best-fit slope, 
sometimes even with an errorbar.  Taken en ensemble,
however, these forced power laws present a very different picture of the IMF.  
Figure 1 is a plot inspired by similar graphs presented by Scalo (1998) and
Kroupa (2001) of the slope of the initial mass function $\Gamma$
where $d N(log M) / d logM \propto M^\Gamma$ and $dN(M)/dM \propto M^{-\alpha}$ such
that $\Gamma$ = 1-$\alpha$, shown as a function of $<M>/M_\odot$ where $<M>$ is the average
mass over which a common slope is claimed to apply.  The data are drawn from an
extensive list of IMF primary references appearing over the last decade.  Many different
techniques are assimilated, including spectroscopic investigations, color-magnitude
diagrams, and luminosity functions.  Similar plots of
$\Gamma$ as a function of minimum mass or maximum mass, 
rather than the average mass, are also instructive.   
What is clear from these Figures
is that at the high mass end (M$>$2-3 $M_\odot$) reported IMF slopes are constant
as a function of mass and distributed
around the original Salpeter value of $\Gamma=-1.35$.
At lower masses (M$<$1-2 $M_\odot$), however, the reported slopes deviate from Salpeter
in a manner such that the IMF slope continually becomes shallower / flatter towards lower
masses.

\begin{figure}
\centering
\includegraphics[height=8cm]{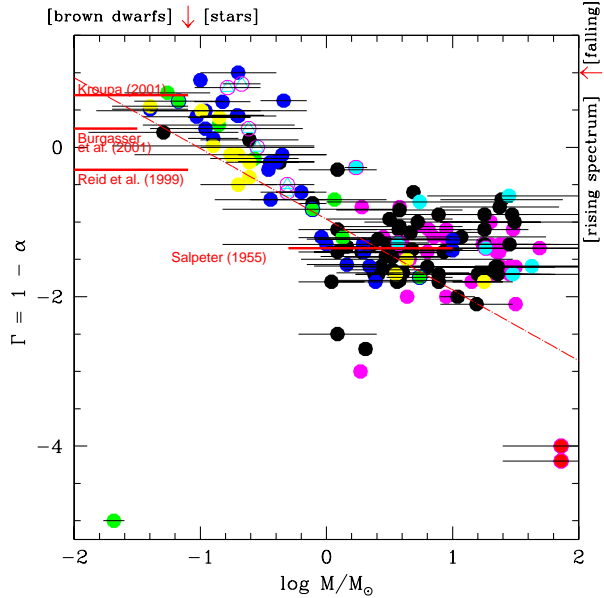}
%\picplace{5cm}{2cm} % Give the correct figure height and width in cm
\caption{
Values quoted in the literature for the slope of the initial mass function 
plotted against mean mass (points), with the range of masses over 
which the single slope is claimed to apply indicated (horizontal lines).
Filled circles represent star-forming regions and young open clusters
while triangles are globular clusters.  Continual change of the mass function 
slope with mean mass below about 1 M$_\odot$
demonstrates that a quadratic or log-normal function best
describes the composite data at low masses.  Above several M$_\odot$
the scatter is more considerable, and the quadratic or log-normal function
yields to a flatter slope that averages close to the Salpeter (1955) value. 
The dash-dot line shows a Miller-Scalo (1979) log-normal.
}
\label{fig:1}       % Give a unique label
\end{figure}

There scatter in Figure 1 amounts to $\sim$1 dex for any mean mass.  Although
it may be tempting to interpret this scatter as real variation in the IMF,
there are several affects which may in part explain it.  As stated
above there are differences in methodology between authors, and while spectroscopic
investigations are indeed the most robust, they are not practical for 
all regions; of note is that there are no systematic differences in Figure 1
between the spectroscopic investigations and those
that are purely photometric (color-magnitude diagrams or luminosity functions). 
Another possible effect is that of systematics present in the different effective temperature
or bolometric correction scales adopted by different authors, but again, offsets between
authors or groups are not evident in the results as presented in the format
of Figure 1.   The adopted set of pre-main sequence theory is
yet another way to introduce systematic differences.  
Finally, the statistical robustness of the results should be 
considered in lieu of generally small sample sizes, particularly at the high mass end
where by the very nature of the IMF the poisson statistics are the poorest.  Kroupa (2001)
has investigated this last point by simulating a version of Figure 1 using 
model populations differing in total numbers of stars (10$^2$-10$^6$) 
sampled many times.  The
observed scatter about the Salpeter slope can in fact be reproduced through sampling 
error alone.  Kroupa (2001) also investigates the scatter introduced by dynamical
evolution of clusters on time scales of a few Myr to tens of Myr, and concludes that
some of the scatter may be attributed to the redistribution of mass within or exterior
to the gravitational bounds of a cluster.

In summary, the general trends emerging from the literature are for a universal IMF
with a single Salpeter-like slope above masses of 1-3 $M_\odot$ and a constantly 
flattening slope towards lower masses, extending smoothly and continuously
from low mass stars into the sub-stellar regime. At present, there is little evidence 
for substantial variations in the IMF from region to region, or over time.  The scatter
present in Figures of IMF slope vs mass are explainable by a combination of methodology
differences, theory differences, and sampling statistics.
Yes, there are outliers in such plots, and the regions in question should
be investigated further to determine the accuracy of the reported results.  But the
dominant trends remain.

\section{Future Goals of IMF Studies}

Having crossed the sub-stellar boundary over the past few years,
the next major threshold
for low mass IMF studies to reach is the theoretical limit for
fragmentation out of a molecular cloud, derived by calculating the ability 
of a Jeans' mass / radius object contracting on
the free-fall time to radiate its own gravitational energy.
Thus far no low-mass cut-off in the IMF has been observed.
Observationally, the Next Generation Space Telescope (now JWST) should be 
sensitive at the distance and age of the Orion Nebula Cluster, for example,
to objects as low in mass as 1 M$_{Jupiter}$.   Given an appropriate means 
of sorting between field star contaminants and cluster members 
at such faint magnitudes, the goal of determining the lowest free-forming mass 
object should be reached in this dense and populous star forming region.

Another major question concerns the universality of the IMF.  Is the mass 
spectrum the same everywhere, that is, invariant with respect to time, location,
size or density of the star-forming environment, metallicity, other initial 
conditions such as temperature or pressure?    Can we relate the IMF 
to some physical aspect 
of the star formation process?  Thus far no systematic variations 
have been detected, for example between high-mass star forming environments 
of varying metallicity (e.g. Massey et al. 1995ab) or between low-mass 
star forming environments of different stellar density (e.g. Luhman 2000), 
or even in ``extreme" environments such as R 136 (Massey \& Hunter 1998) 
in the Large Magellenic Cloud.   That is not to say
that no such trends are present.  Rather, to within our ability to provide robust
statistical assessment, variations have not yet been detected.
A very few regions with ``unusual" IMFs have been presented (e.g. the Hillenbrand
et al. 1995 discussion of the bias of the BD+40 4124 region towards higher 
mass stars and the Briceno et al. 2002 presentation of
the lack of brown dwarfs in Taurus) though whether these regions 
are truly unusual or merely the expected statistical outliers remains unquantified.
A goal then is to more rigorously quantify the invariance of the IMF.

\section{Summary}

\begin{itemize}
\item[$\bullet$]
The IMF is not a single power law but rather continually changes its slope
below masses of 2-3 M$_\odot$.

\item[$\bullet$]
There is nothing special about the hydrogen burning limit in terms of the 
frequency distribution of masses above or below this boundary.

\item[$\bullet$]
Thus far no low-mass cut off has been observed in the IMF, 
with limits reaching perhaps as low as 20-30 M$_{Jupiter}$ 
in some star formation regions.  The next
fundamental boundary is that of opacity-limited fragmentation.

%\ite[{$\bullet$]
%Some uncertainty remains concerning the high mass cut-off.  The highest mass stars
%observed are $120-130 M_\odot$ with the Eddington limit expected to be closer to
%$180 M_\odot$.   Given all the high mass star forming regions we know about, should we
%have expected to see a 180 M$_\odot$ star by now?

\item[$\bullet$]
The lack of strong evidence for IMF variations suggest that its functional form may
be universal and therefore independent of any of the physical variables suspected to
play a role in the star formation process.
%(metallicity, density, temperature, turbulence, etc.).   

\end{itemize}

%%
% BibTeX users please use
% \bibliographystyle{}
% \bibliography{}
%
% Non-BibTeX users please follow the syntax
% the syntax of "referenc.tex" for your own citations
%%%%%%%%%%%%%%%%%%%%%%%% referenc.tex %%%%%%%%%%%%%%%%%%%%%%%%%%%%%%
% sample references
% "physics"
%
% Use this file as a template for your own input.
%
%%%%%%%%%%%%%%%%%%%%%%%% Springer-Verlag %%%%%%%%%%%%%%%%%%%%%%%%%%

%
% BibTeX users please use
% \bibliographystyle{}
% \bibliography{}

%%%%%%%%%%%%%%%%%%%%%%%%%%%%%%%%%%%%%%%%%%%%%%%%%%%%%%%%%%%%%%%%%%%%%%  }

%%%%%%%%%%%%%%%%%%%%%%%%%%%%%%%%%%%%%%%%%%%%%%%%%%%%%%%%%%%%%%%%%%%%%%

\printindex
\end{document}